\documentclass[a4paper,twoside]{article}

\usepackage{epsfig}
\usepackage{subcaption}
\usepackage{calc}
\usepackage{amssymb}
\usepackage{amstext}
\usepackage{amsmath}
\usepackage{amsthm}
\usepackage{multicol}
\usepackage{pslatex}
\usepackage{natbib}
\usepackage{cleveref}
\usepackage{tikz}
\usetikzlibrary{calc}
\usepackage{pgfplots}
\pgfplotsset{compat=newest}
\usepgfplotslibrary{statistics}
\usepgfplotslibrary{groupplots}
\usepackage{siunitx}
\usepackage{bm}
\usepackage{stmaryrd}
\usepackage{graphicx}  
\usepackage{mathtools}
\usepackage{enumitem}
\usepackage{booktabs}
\usepackage[utf8]{inputenc}

\usepackage{SCITEPRESS}     

\definecolor{clr1}{RGB}{57,106,177}
\definecolor{clr2}{RGB}{218,124,48}
\definecolor{clr3}{RGB}{62,150,81}
\definecolor{clr4}{RGB}{204,37,41}

\newcommand{\cels}[1]{\SI{#1}{\celsius}}

\newcommand{\kW}[1]{\SI{#1}{\kilo\watt}}
\newcommand{\kWh}[1]{\SI{#1}{\kilo\watt\hour}}
\newcommand{\mins}[1]{\SI{#1}{\minute}}
\newcommand{\secs}[1]{\SI{#1}{\second}}
\newcommand{\hrs}[1]{\SI{#1}{\hour}}
\newcommand{\percents}[1]{\SI{#1}{\percent}}

\newcommand{\abbrMILP}{MILP} 

\newcommand{\idxJob}{i} 
\newcommand{\idxJobAnother}{j} 
\newcommand{\idxMach}{k} 

\newcommand{\symbJob}{J} 
\newcommand{\symbMach}{M} 
\newcommand{\symbJobDS}{\symbJob_{s}} 
\newcommand{\symbJobDE}{\symbJob_{e}} 
\newcommand{\symbStart}{s} 
\newcommand{\symbAss}{a} 
\newcommand{\symbAssVec}{\bm{\symbAss}} 
\newcommand{\symbAssElem}[1]{\symbAss_{#1}} 
\newcommand{\symbMode}{v} 
\newcommand{\symbOn}{\text{proc}} 

\newcommand{\cJobs}{n} 
\newcommand{\cMach}{m} 
\newcommand{\cP}[1]{p_{#1}} 
\newcommand{\cR}[1]{r_{#1}} 
\newcommand{\cD}[1]{\tilde{d}_{#1}} 
\newcommand{\cBigM}{\text{\rmfamily\upshape M}} 

\newcommand{\cMode}[1]{P_{#1}} 
\newcommand{\cSwitchT}[1]{T_{#1}} 
\newcommand{\cSwitchC}[1]{E_{#1}} 
\newcommand{\cCostOnOff}{E^{(\text{start})}} 
\newcommand{\cProcJob}[1]{E^{(\text{proc})}_{#1}} 
\newcommand{\cHorizon}{H} 

\newcommand{\cToper}{T_{\text{oper}}} 

\newcommand{\cJob}[1]{\symbJob_{#1}} 

\newcommand{\setModes}{\mathcal{V}} 
\newcommand{\setJobs}{\mathcal{\symbJob}} 
\newcommand{\setJobsAll}{\{ \symbJob_1, \symbJob_2, \dots, \symbJob_{\cJobs} \}} 
\newcommand{\setJobsS}{\setJobs \cup \{ \symbJobDS \}} 
\newcommand{\setJobsE}{\setJobs \cup \{ \symbJobDE \}} 
\newcommand{\setMach}{\mathcal{\symbMach}} 
\newcommand{\setMachAll}{\{\symbMach_1, \symbMach_2, \dots, \symbMach_{\cMach} \}} 

\newcommand{\setNonNegInt}{\mathbb{Z}_{\ge0}} 
\newcommand{\setPosReal}{\mathbb{R}_{>0}} 
\newcommand{\setNonNegReal}{\mathbb{R}_{\ge0}} 

\newcommand{\sol}{S} 
\newcommand{\solSVec}{\bm{\symbStart}} 
\newcommand{\solS}[1]{\symbStart_{#1}} 
\newcommand{\solAssVec}{\bm{\symbAss}} 
\newcommand{\solAss}[1]{\symbAss_{#1}} 

\newcommand{\vJobAss}[2]{a_{#1,#2}} 
\newcommand{\vJobStart}[1]{s_{#1}} 
\newcommand{\vJobPred}[3]{y_{#1,#2,#3}} 
\newcommand{\vGap}[2]{z_{#1,#2}} 
\newcommand{\vMachOn}[1]{x_{#1}} 

\DeclareMathOperator{\pred}{pred_{\sol}}

\newcommand{\fESymb}{E} 
\newcommand{\fE}[1]{\fESymb(#1)} 
\newcommand{\fEx}[1]{\mathbb{E}(#1)} 
\newcommand{\fUniform}[2]{\mathit{U}\{#1,#2\}} 
\newcommand{\fExponential}[1]{\mathit{Exp}(#1)} 

\usetikzlibrary{pgfplots.external}
\tikzset{external/system call={pdflatex \tikzexternalcheckshellescape 
                                        -halt-on-error
                                        -interaction=batchmode 
                                        -jobname "\image" "\texsource"
                                        && pdftops -eps "\image.pdf"}}
\begin{document}

\title{On Idle Energy Consumption Minimization in Production: \\ Industrial Example and Mathematical Model}

\author{\authorname{Ond\v{r}ej Benedikt\sup{1,2}\orcidAuthor{0000-0002-7365-844X},
P\v{r}emysl \v{S}\r{u}cha\sup{1}\orcidAuthor{0000-0003-4895-157X}
and Zden\v{e}k Hanz\'{a}lek\sup{1}\orcidAuthor{0000-0002-8135-1296}
}
\affiliation{\sup{1}Czech Institute of Informatics, Robotics and Cybernetics, Czech Technical University in Prague,\\ Jugosl\'{a}vsk\'{y}ch partyz\'{a}n\r{u} 1580/3, Prague,
Czech republic}
\affiliation{\sup{2}Czech Technical University in Prague, 
Faculty of Electrical Engineering, 
Department of Control Engineering, 
Karlovo~n\'{a}m\v{e}st\'{i} 13, Prague, Czech republic}
\email{\{ondrej.benedikt, premysl.sucha, zdenek.hanzalek\}@cvut.cz}
}


\abstract{This paper, inspired by a real production process of steel hardening, investigates a scheduling problem to minimize the idle energy consumption of machines.
The energy minimization is achieved by switching a machine to some power-saving mode when it is idle.
For the steel hardening process, the mode of the machine (i.e., furnace) can be associated with its inner temperature. Contrary to the recent methods, which consider only a small number of machine modes, the temperature in the furnace can be changed continuously, and so an infinite number of the power-saving modes must be considered to achieve the highest possible savings. To model the machine modes efficiently, we use the concept of the energy function, which was originally introduced in the domain of embedded systems but has yet to take roots in the domain of production research. The energy function is illustrated with several application examples from the literature. Afterward, it is integrated into a mathematical model of a scheduling problem with parallel identical machines and jobs characterized by release times, deadlines, and processing times.
Numerical experiments show that the proposed model outperforms a reference model adapted from the literature.}

\onecolumn \maketitle \normalsize \setcounter{footnote}{0} \vfill

\section{INTRODUCTION}

    \noindent In recent years, there has been an increasing interest in energy-efficient scheduling \citep{2016_gahm,2019:Gao}. The reasons are both ecological and economical. By implementing efficient scheduling, a significant amount of energy can be saved \citep{2007_mouzon,2016_gahm} with negligible investments.
    
\begin{figure}
        \centering
        \includegraphics[width=\columnwidth]{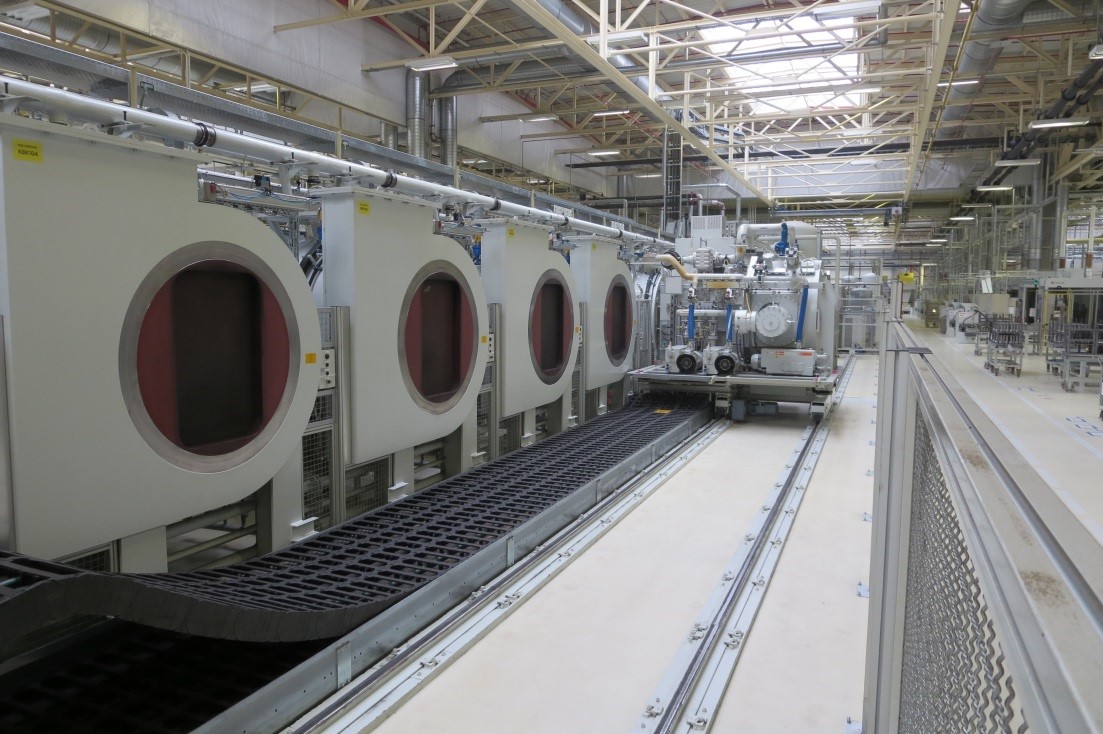}
        \caption{A steel hardening production line in Škoda Auto consisting of electrical vacuum furnaces \citep{2016:Dusek-thesis}.}
        \label{fig:furnaces}
    \end{figure}

    This paper addresses a scheduling problem to minimize the total idle energy consumption of the machines.
    Our work is inspired by a steel hardening process, which has high energy demands. During the process, a material is heated to a very high temperature (defined by the technological process) in one of the identical furnaces. A segment of the production line with several furnaces in Škoda Auto company is shown in \Cref{fig:furnaces}. There, parts of the future gear shafts are hardened. The operating temperature of the furnaces is \cels{960}. Typically, the furnaces are heated to the operating temperature at the beginning of the weak, and cooled down at its end. However, some of the furnaces might not be utilized all the time, depending on the previous production stages. 
    
\begin{figure}
    \centering
    \includegraphics{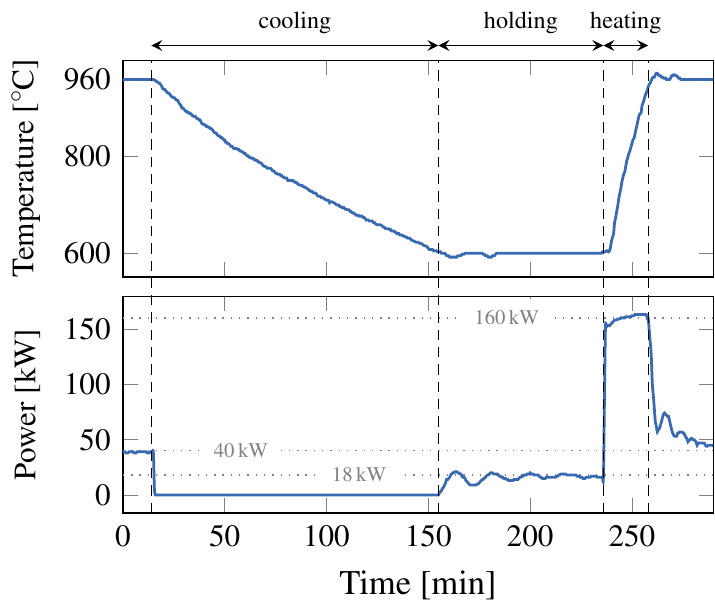}
    \caption{A relationship between the power consumption and the temperature of a single furnace during the experiment.}
    \label{fig:power-and-temperature}
\end{figure}

    If the furnace is underutilized, energy can be saved by lowering the temperature during the idle periods. However, the furnace needs to be heated back to the operating temperature before the next job arrives. As re-heating of the furnace takes time and consumes energy, it needs to be planned carefully. Our preliminary study in Škoda Auto company has shown
    that over \percents{5} of the energy consumption of the hardening line could be saved by switching the idle furnaces to the power-saving modes \citep{2016:Dusek-thesis}.
    Several experiments were performed to investigate the potential for energy savings. During one of the experiments, the furnace was heated to the operating temperature, i.e., \cels{960}; afterward, it cooled to \cels{600} and was re-heated back to \cels{960}. \Cref{fig:power-and-temperature}
    shows the relationship between the temperature and power. Note that during the cooling phase, the input power is zero. On the other hand, the maximal power (\kW{160}) is applied for the re-heating to reach the operating mode as soon as possible. The experiment shows that the power needed to compensate for the losses when holding \cels{960} (about \kW{40}) is more than two times higher than the power needed for \cels{600} (about \kW{18}). That indicates the potential for energy saving. Of course, if the machine were turned off completely (i.e., to the ambient temperature), the power consumption during the holding phase would be zero. However, the complete cooling and re-heating would require a considerably longer time (which might not be available). Also, the energy needed for re-heating would be higher. Therefore, depending on the idle period length, different savings might be achieved by cooling to different temperatures.
    
    \subsection{Contributions and Outline}
    
    This study aims to contribute to the growing area of energy optimization research by providing an example of the parallel machine problem, its formal description, and a formulation of a Mixed Integer Linear Programming (\abbrMILP{}) model. Compared to the other problems of the idle energy optimization studied in the literature, this one is special because the slow dynamics of the machine (furnace) and its ``inner state'' (temperature) cannot be neglected. 
    The proposed \abbrMILP{} model takes into account possible power-saving modes indirectly by using the energy function, which abstracts the dynamics of the machine.
    The proposed model is compared with another \abbrMILP{} model, adapted from the literature, and its performance is further tested on a set of benchmark instances.
    Also, this paper shows that the energy function can easily replace conventional modeling approaches (e.g., those using on/off or on/idle/off modes).

    The rest of the paper is organized as follows. \Cref{sec:related-work} summarizes connections to relevant literature. \Cref{sec:problem-statement} formally describes the studied scheduling problem. Afterward, the abstraction used for idle energy optimization, called the \textit{energy function}, is introduced in \Cref{sec:energy-function}, together with several examples from the literature. \Cref{sec:model} describes the \abbrMILP{} model, which integrates the energy function. In \Cref{sec:experiments}, the proposed model is compared to a reference model, and its scalability is tested on a range of instances. Finally, \Cref{sec:conclusions} gives a summary of the findings.

\section{RELATED WORK} \label{sec:related-work}

\noindent In the area of production scheduling, one of the first analyses of underutilized machines was done by \citet{2007_mouzon}, who observed that changing the machine modes could reduce energy consumption significantly. Furthermore, they proposed a mathematical model to optimize energy and the completion time by turning the machine on and off. However, the model optimized a single machine only, assuming a fixed order of the jobs.

In the domain of single machine problems, \citet{2014_shrouf} proposed a mathematical model and heuristics for a single machine with three modes (processing, idle and shut down), assuming variable energy prices. \citet{2016_gong} also addressed a problem with dynamic energy pricing, modeling a single machine with multiple processing modes by a finite state automaton. After completion of each job, the machine could stay idle, shut down, or start processing the next job. No power-saving modes besides the shutdown mode were allowed. Furthermore, it was assumed that transitions between the individual modes of the machine are immediate. This assumption cannot be used here, because heating and cooling take time. A single machine problem was also investigated by \citet{2017_che}, who optimized weighted energy consumption and makespan and deployed cluster analysis to approximate Pareto fronts of large-size problems. Their problem statement is similar to ours, but they scheduled on and off modes only, which might be too restrictive when the complete power-down takes a long time. 

Compared to the single machine problem, parallel machine scheduling studied here is much harder, because the assignment of the jobs to machines needs to be found together with their relative order and start times. \citet{2013_fang} described a \abbrMILP{} model, dispatching heuristics, and a particle swarm optimization algorithm for the weighted tardiness/cost problem. They assumed that the processing time of a job could be shortened by increasing the speed of the machine, which is not the case for the problem studied here  because the hardening process needs to follow a given technological specification. Furthermore, their \abbrMILP{} model was relatively slow even for five jobs scheduled on three machines. \citet{2015_liang} formulated a non-linear model of an on/off problem with unrelated parallel machines and proposed an ant optimization heuristic. \citet{2017_masmoudi} developed a mathematical model and heuristic for a flow shop problem with time-of-use energy pricing, and \citet{2019:Meng} proposed \abbrMILP{} models for a flexible job shop problem, where on and off modes of the machines were considered. \citet{2018_benedikt} formulated a parallel machine scheduling problem with multiple processing modes with different processing speeds and power consumption. A decomposition-based approach based on the column generation technique was proposed to solve the problem. However, the resulting mathematical models are still very complex and have limited scalability. 
\cite{2019:Modos} studied a scheduling problem with dedicated machines, inspired by glass tempering and steel hardening processes. Instead of energy minimization, the energy consumption limit (imposed for each metering interval) was considered. However, the authors did not model the power-saving modes of the machine, which might have a positive impact on energy consumption in the metering intervals when some of the machines are idling.

Reviews of energy-efficient scheduling were published by \citet{2016_gahm} and \citet{2019:Gao}, who summarized the latest trends in the area. The reviews show that interest in energy-efficient scheduling is steadily increasing year by year. Also, the reviews show that a mixed-integer (linear) programming is one of the standard modeling techniques used in the area.

We conclude that although the idle energy minimization is widely studied, only two or several modes of the machines are commonly modeled, which might be too restrictive. Also, the unifying concept of the idle energy consumption modeling (i.e., the energy function) is still not widely known.

\section{PROBLEM STATEMENT} \label{sec:problem-statement}

\subsection{Input Parameters and Assumptions}

We consider a finite set of jobs $\setJobs = \setJobsAll$ and a finite set of parallel identical machines $\setMach = \setMachAll$. Each job $\cJob{\idxJob}$ is characterized by three non-negative integers: processing time $\cP{\idxJob}$, release time $\cR{\idxJob}$, and deadline $\cD{\idxJob}$. Release time and deadline form an execution time window within which the job needs to be processed. Each job can be processed on any machine, but each machine can process only a single job at a time. When the processing starts, it cannot be preempted.

While a job is processed, the machine needs to be operating in a \emph{processing mode} (heated to predefined operational temperature $\cToper$)  and cannot change it until the processing of the job is finished. The processing of job $\cJob{\idxJob}$ consumes energy $\cProcJob{\idxJob}$. When a machine does not process any job, it enters a so-called \emph{idle period}. During the idle period, the temperature can be changed to achieve energy savings, but when the next job arrives, the machine needs to be heated back to $\cToper$. The relationship between energy consumption and the length of the idle period $\Delta$ is given by energy function $\fESymb: \setNonNegReal \rightarrow \setNonNegReal$. For  fixed $\Delta$, value $\fE{\Delta}$ represents the best attainable energy consumption across all available machine modes. The energy function is further discussed in \Cref{sec:energy-function}.

It is assumed that each machine starts and ends in ``off-mode'', which has zero power consumption. Some machines may remain in the off-mode all the time, but at least one machine needs to be turned on to process the jobs. The energy needed for the switching from off-mode to the processing mode and back is given by $\cCostOnOff \in \setPosReal$. It is assumed that there is enough time to turn the machines on before the first job is available and to turn them off after the last job is processed. In the following text, the length of the scheduling horizon is denoted by $\cHorizon$, where $\cHorizon = \max \{\cD{\idxJob} \mid \symbJob_{\idxJob} \in \setJobs \}$.

\subsection{Solution Representation}

Solution $\sol$ can be represented by a pair of vectors, $\sol = (\solAssVec,\solSVec)$, where vector \hbox{$\solSVec = (\solS{1}, \dots, \solS{\cJobs}) \in \setNonNegInt^{\cJobs}$} represents start times of the jobs, and vector \hbox{$\solAssVec = (\solAss{1}, \dots, \solAss{\cJobs}) \in \setMach^{\cJobs}$} captures assignment of the jobs to machines. 
The solution is feasible if all jobs are processed within their execution windows without preemption or overlapping.

\subsection{Optimization Objective} \label{sec:objective}

The assignment of the jobs, together with their start times, define the idle periods in the schedule. Depending on their lengths, energy consumption may vary.

Before defining the objective, let us denote the index of the job scheduled immediately before job $\cJob{\idxJob}$ in solution $\sol$ on machine $\solAss{\idxJob}$ by $\pred(\idxJob)$. If such a job does not exist, i.e., the job $\cJob{\idxJob}$ is the first on machine $\solAss{\idxJob}$, we define $\pred(\idxJob) := 0$. Now, the energy consumption corresponding to solution $\sol$ can be written as follows:

\begin{small}

\begin{equation} \label{eq:sol-obj}
    \sum\limits_{\idxJob=1}^{\cJobs} \cProcJob{\idxJob} + \sum\limits_{\mathclap{\substack{\idxJob = 1\\ \pred(\idxJob) \neq 0}}}^{\cJobs} \fE{\solS{\idxJob} - \solS{\pred(\idxJob)} - \cP{\pred(\idxJob)}} + \sum\limits_{\mathclap{\substack{\idxJob = 1\\ \pred(\idxJob) = 0}}}^{\cJobs} \cCostOnOff,
\end{equation}

\end{small}

where the first sum represents the total energy needed to process the jobs, the second sum corresponds to the energy consumed during the idle periods (energy function $\fESymb$ is further described in \Cref{sec:energy-function}), and the third sum express the energy needed for turning the machines (to which at least one job is assigned) \emph{on} and \emph{off}. The \emph{optimal schedule} minimizes the total energy consumption \eqref{eq:sol-obj}.

Note that constant $\sum_{\idxJob=1}^{\cJobs} \cProcJob{\idxJob}$ can be omitted for the optimization as it does not affect the structure of the optimal schedule. 

\subsection{Complexity}

Using the standard $\alpha\vert\beta\vert\gamma$ notation, we can characterize the studied problem as $P \vert r_j, \tilde{d}_j \vert E_{\text{idle}}$, meaning idle energy minimization for parallel identical machines and jobs characterized by release times and deadlines.
The problem is $\mathcal{NP}$-hard because its subproblem $1\vert r_j, \tilde{d}_j \vert -$ is already $\mathcal{NP}$-complete in the strong sense \citep{1977:Garey}.

\section{ENERGY FUNCTION} \label{sec:energy-function}

\noindent The energy function is a concept used when idle energy minimization is taken into account. The specific properties of the resource are abstracted, and only the information about the energy consumption is explicitly represented. This concept often simplifies the analysis of the problem properties, as well as the models and algorithms. Here, we establish the basic notions in \Cref{sec:en-basics} and show several examples from the literature in \Cref{sec:en-examples}. Afterward, we discuss the energy function obtained for vacuum furnaces studied in this paper in \Cref{sec:en-furnace}.

\subsection{Basic Notions} \label{sec:en-basics}

Denoting $\setModes$ the set of all machine modes, a basic machine model can be described as follows. When the machine processes a job, it needs to be operating in the processing mode ``$\symbOn$''. When no job is processed, the machine can change the mode. Switching to (non-processing) mode $\symbMode$ and back takes time $\cSwitchT{\symbMode}$, and consumes energy $\cSwitchC{\symbMode}$. The power consumption of the machine in mode $\symbMode$ is denoted by $\cMode{\symbMode}$, and is assumed to be constant. 
Depending on the length of the idle period between two neighboring jobs, which is denoted by $\Delta$, a machine can either turn to some non-processing mode (and back) or remain in the processing mode. Clearly, if $\Delta$ is shorter than $\cSwitchT{\symbMode}$, switching to mode $\symbMode$ is not feasible. Also, switching to a power-saving mode should be only performed when it would not increase the overall power consumption \citep{2012_devadas}.
These observations lead us to the following formulation of the energy consumption function $\fESymb: \setNonNegReal \rightarrow \setNonNegReal$ \citep{2013_gerards}:

\begin{equation} \label{eq:energy-consumption}
   \fE{\Delta} = \min_{\symbMode \in \setModes: \Delta \geq \cSwitchT{\symbMode}} \{ \cSwitchC{\symbMode} + \cMode{\symbMode} \cdot (\Delta - \cSwitchT{\symbMode}) \}.
\end{equation}

\noindent Formally, we define $\cSwitchT{\symbOn} = \cSwitchC{\symbOn} = 0$, i.e., for \hbox{$\setModes = \{ \symbOn \}$}, $\fE{\Delta} = \cMode{\symbOn} \cdot \Delta$.

The value $\fE{\Delta}$ represents the best attainable energy consumption for given idle period length $\Delta$. Note that the optimal mode to which the machine should be switched is given by the argument of the minimum in \eqref{eq:energy-consumption}.
In the following section, we discuss several applications of this model.

\subsection{Examples From the Literature} \label{sec:en-examples}

Concept of the idle energy minimization is relatively old. Initially, the possibilities of the power savings were investigated in the field of embedded systems \citep{2000_Benini,2008_augustine,2013_gerards}, where energy consumption plays a critical role for the lifetime of the battery-powered systems. Nowadays, it is common that hardware components have one or several power-saving modes defined by the manufacturer \citep{2013_gerards}. Of course, the time and energy needed to perform the transitions between the modes are usually not negligible. Therefore, the selection of the ``optimal'' power-saving mode of the component depends on the idle period length.

\begin{figure}
    \centering
    \includegraphics{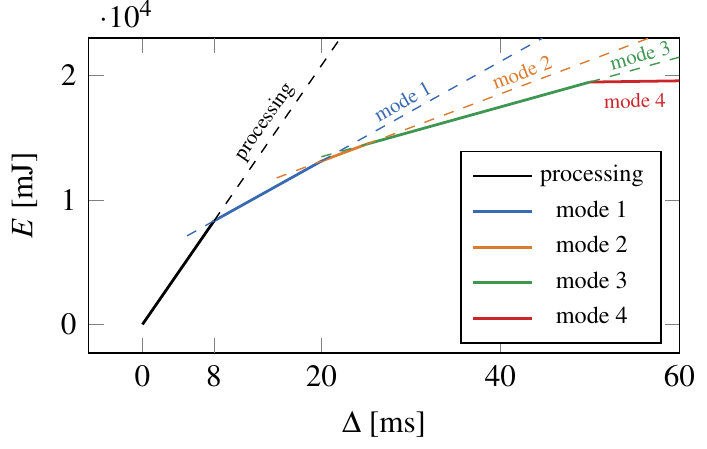}
    \caption{The energy function of a sensor node from embedded systems domain \citep{2001_sinha}.}
    \label{fig:sensor-node}
\end{figure}

An example of a hardware device with multiple modes is a sensor node \citep{2001_sinha}, which has four power-saving modes, with $\cSwitchT{\symbMode}$ equal to 5, 15, 20, and 50 milliseconds, respectively. The piecewise linear function drawn by a solid line in \Cref{fig:sensor-node} is the energy function of the device. It consists of 5 segments, corresponding to different modes of the device. Note that even though mode 1 is accessible already for $\Delta = 5$, it is not profitable to perform the switching until $\Delta = 8$, which is called a \emph{break-even time} in the literature \citep{2012_devadas}.

\begin{figure}[b]
    \centering
    \resizebox{\columnwidth}{!}{%
    \includegraphics{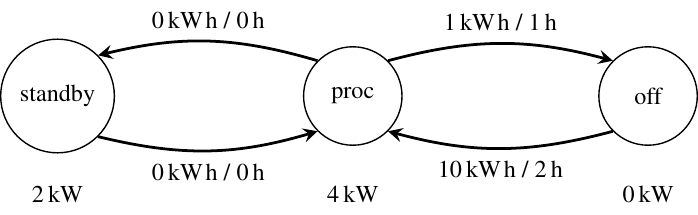}
    }
    \caption{Example of a transition graph depicting machine modes and transitions between them \citep{2014_shrouf}.}
    \label{fig:ex-trans-graph} \vspace{10pt}
    
    \includegraphics{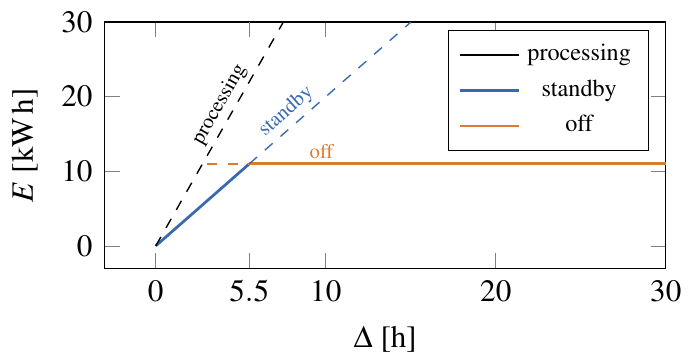}
    \caption{Energy function corresponding to the transition graph depicted in \Cref{fig:ex-trans-graph}.} 
    \label{fig:ex-energy-func} 
\end{figure}

In the production domain, research of the idle energy minimization is still relatively new. The interest in the topic increased after Mouzon's study, which analyzed underutilized machines \citep{2007_mouzon}. Since that time, researchers have started to take the power-saving modes of the machines into consideration. However, only a small number of modes is usually considered \citep{2014_shrouf,2017_che,2019_Aghelinejad}. Typically, the machine modes and transitions are modeled by a transition graph. A representative example of such a graph is shown in \Cref{fig:ex-trans-graph} \citep{2014_shrouf}. The nodes are labeled by power consumption, while the edges are labeled by energy/time needed for the transition. Parameters needed for the energy function can be easily obtained from the graph. For example, $\cSwitchT{\text{off}} = \hrs{3}$, $\cSwitchC{\text{off}} = \kWh{11}$, $\cMode{\text{off}} = \kW{0}$, etc. The corresponding energy function is shown in \Cref{fig:ex-energy-func}.

\subsection{Energy Function and Industrial Furnaces} \label{sec:en-furnace}

When considering the industrial furnaces, the situation is slightly more complicated because of their slow dynamics. Depending on the input power, the furnace is either cooling (temperature inside is decreasing), holding (temperature inside is stable), or heating (temperature inside is increasing), as shown in \Cref{fig:power-and-temperature}. We could associate the mode $\symbMode$ of the furnace with the temperature that should be held inside of the furnace. The power consumption $\cMode{\symbMode}$ is then the power needed to compensate for the steady-state losses of the furnace. The switching time $\cSwitchT{\symbMode}$ is the time needed for cooling plus the time needed for re-heating, and the switching energy consumption $\cSwitchC{\symbMode}$ is equal to the energy needed for cooling (which equals zero) and the energy needed for re-heating.

Now, if we wanted to model such a system by a transition graph, we would face a problem, as only a finite number of machine modes can be modeled like that. On the other hand, the temperature in the furnace can be changed continuously, and so an infinitely large set $\setModes$ would be needed to model all possible power-saving modes. 

\begin{figure}
    \centering
    \includegraphics{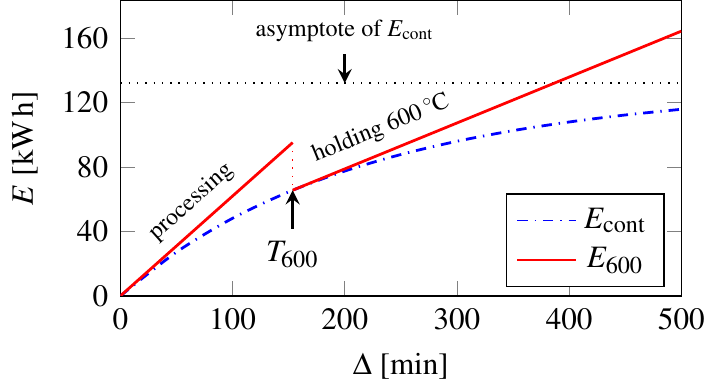}
    \caption{Energy function of an industrial furnace $\fESymb_{\text{cont}}$ integrating all power-saving modes  \citep{2019:Benedikt}, and function $\fESymb_{\text{600}}$ corresponding to a single power-saving mode.}
    \label{fig:energy-furnace}
\end{figure}

\Cref{fig:energy-furnace} shows two possible energy functions of the furnace corresponding to two different control strategies. Function $\fESymb_{600}$ represents a standard approach using a transition graph with a single power-saving mode associated with the temperature \cels{600}. The corresponding control rule states: cool to \cels{600}, hold \cels{600} as long as possible and re-heat to the operating temperature as fast as possible just before the end of the idle period. The slope of the first segment of $\fESymb_{600}$ corresponds to the power consumption of the processing mode, while the slope of the second segment corresponds to the power compensating for the losses when holding \cels{600}. Energy $\fESymb_{600}(\cSwitchT{600})$ is the energy needed to re-heat the furnace from \cels{600} back to the \cels{960} (remember that energy needed for cooling is zero). Transition to the power-saving mode is not possible during time interval $[0,\cSwitchT{600})$, because there is not enough time to cool the furnace to \cels{600} and re-heat it back.

On the other hand, function $\fESymb_{\text{cont}}$ is a result of a detailed analysis of the industrial furnace, see \cite{2019:Benedikt}. The control rule states that the furnace should be cooling as long as possible, and should be re-heated as fast as possible to reach the operating mode again at the end of the idle period. With the increasing length of the idle period, the temperature to which the furnace cools decreases until it reaches the ambient temperature. Then, the energy needed to cool to the ambient temperature and to re-heat back corresponds to the asymptote shown by the dashed line.

Note that when only a subset of all modes is considered, the energy function becomes discontinuous, like $\fESymb_{600}$. The reason for the discontinuity is that the cooling of the furnace is slow, and so the integral of the processing power over interval $[0,\cSwitchT{600}]$ becomes greater than the switching energy consumption $\cSwitchC{600}$. Contrary to that, function $\fESymb_{\text{cont}}$, which implicitly represents all possible temperatures, remains continuous and concave (similarly to the energy functions shown in Figures \ref{fig:sensor-node} and \ref{fig:ex-energy-func}). 

\section{MATHEMATICAL MODEL} \label{sec:model}

\noindent In this section, we integrate the energy function into a \abbrMILP{} model proposed for the problem defined in \Cref{sec:problem-statement}.

As shown by \cite{2016_gahm}, formulating problems by \abbrMILP{} models and solving them by standard solvers has been one of the widely used approaches to the optimal energy-aware scheduling. There exist several alternative approaches to the scheduling problem modeling in \abbrMILP{}. Many authors use time-indexed models \citep{2014_shrouf,2012_mitra,2017_masmoudi} -- scheduling horizon is divided into periods and decisions about the mode of the machine or assignment of jobs have to be made for each period separately. In many cases, the time-indexed models are the only reasonable alternative, e.g., when the properties of the system vary through time. However, the size of a time-indexed model depends on the length of the scheduling horizon, which is prohibitive.

Other formulation approaches use event-based modeling \citep{2011_kone}, permutation (position-based) models \citep{2017_che}, or relative order models \citep{2015_liang}. 

The model we propose in this work is a special variant of the relative-order model, modeling the direct predecessor and successor of each job. A similar idea has already been used in the scheduling domain successfully \citep{2008_liu}. However, we integrate the concept of the energy consumption function into the model formulation to describe the transition costs efficiently. The model is compared to a position-based model in \Cref{sec:experiments}.

\subsection{Model Description}

At first, we define two dummy jobs; the first one, $\symbJobDS$ (dummy-start), is fixed to start and end at time 0, while the second one, $\symbJobDE$ (dummy-end), must start and end at time $\cHorizon$. Both have zero processing length. These dummy jobs model the predecessor (successor) of the first (last) job assigned to each machine.

\subsubsection{Variables}

We use five types of variables to model the problem. Binary variables are:

\begin{itemize}[leftmargin=2.75em]
  \item[$\vJobAss{\idxJob}{\idxMach}$] indicates if job $\symbJob_{\idxJob}$ is assigned to machine $\symbMach_{\idxMach}$
  \item[$\vJobPred{\idxJob}{\idxJobAnother}{\idxMach}$] decides whether job $\symbJob_{\idxJob}$ immediately precedes job $\symbJob_{\idxJobAnother}$ on machine $\symbMach_{\idxMach}$
  \item[$\vMachOn{\idxMach}$] indicates whether at least one job is assigned to machine $\symbMach_{\idxMach}$
\end{itemize}

\noindent Continuous variables follow:

\begin{itemize}[leftmargin=2.75em]
  \item[$\vJobStart{\idxJob}$] corresponds to start time of job $\symbJob_{\idxJob}$
  \item[$\vGap{\idxJob}{\idxJobAnother}$] equals to the length of the idle period between jobs $\symbJob_{\idxJob}$ and $\symbJob_{\idxJobAnother}$ if $\symbJob_{\idxJob}$ immediately precedes $\symbJob_{\idxJobAnother}$, otherwise 0
\end{itemize}

Variables $\vJobPred{\idxJob}{\idxJobAnother}{\idxMach}$ are defined for jobs including the dummies. All the other variables are defined for non-dummy jobs only. Note that variables $\vGap{\idxJob}{\idxJobAnother}$ model the lengths of idle periods $\Delta$.

\subsubsection{Objective}

As discussed in \Cref{sec:objective}, the objective is to minimize the total idle energy consumption plus the energy needed to power the machines up and turn them down (if there is at least one job assigned to the machine). Using the defined variables, we can write the whole objective as

\begin{equation} \label{eq:model-objective}
    \sum\limits_{\idxJob \in \setJobs} \sum\limits_{\substack{\idxJobAnother \in \setJobs\\ \idxJob \neq \idxJobAnother}}\fE{\vGap{\idxJob}{\idxJobAnother}}
    + \cCostOnOff \cdot \sum\limits_{\idxMach \in \setMach} \vMachOn{\idxMach}. \\
\end{equation}

The non-linear energy function $\fESymb$ is approximated by a piecewise linear function. The approximation can be made simply and with high precision even with a small number of segments, thanks to the simple shape of the energy function.

Note that functions corresponding to finite transitions graph, such as the ones shown in \Cref{fig:sensor-node} and \Cref{fig:ex-energy-func}, are already piecewise linear by definition \eqref{eq:energy-consumption}.

The piecewise linear objective function can be further linearized by introducing additional binary and continuous variables. However, modern solvers, such as Gurobi or CPLEX, can optimize piecewise linear objectives natively. The linearization is handled internally, and solvers may even benefit by using specialized data structures and algorithms.

\subsubsection{Constraints} 

Equations (\ref{eq:model-assignment}) force each job to be assigned to exactly one machine. Constraints (\ref{eq:model-predecessor}) and (\ref{eq:model-successor}) define that if job $\symbJob_{\idxJob}$ is scheduled to machine $\symbMach_{\idxMach}$, it has exactly one immediate predecessor and successor on that machine, and otherwise if it is not assigned to machine $\symbMach_{\idxMach}$, it cannot precede and follow any other job on that machine. Constraints (\ref{eq:model-dummy-s}) and (\ref{eq:model-dummy-e}) force dummy-start (dummy-end) to have exactly one successor (predecessor) on each machine. Execution time windows of the jobs are established by (\ref{eq:model-rdp}), whereas constraints (\ref{eq:model-overlaps}) forbid overlapping of the neighboring jobs. Inequalities (\ref{eq:model-z-UB}), (\ref{eq:model-z-LB}) and (\ref{eq:model-z-zero}) link variable $\vGap{\idxJob}{\idxJobAnother}$ to the length of the respective idle period. If job $\symbJob_{\idxJob}$ precedes job $\symbJob_{\idxJobAnother}$ on some machine, $\vGap{\idxJob}{\idxJobAnother}$ is exactly equal to $\vJobStart{\idxJobAnother} - (\vJobStart{\idxJob} + \cP{\idxJob})$, i.e., to the length of the idle period between those two consecutive jobs; otherwise, it is set to zero. Finally (\ref{eq:model-machine-on}) forces $\vMachOn{\idxMach}$ to 1, if there is at least one job assigned to this machine. Symbol $\cBigM$ represents some large constant (e.g., $\cBigM = \cHorizon$).

\begingroup
\allowdisplaybreaks

\begin{align}
    & \sum\limits_{\symbMach_{\idxMach} \in \setMach} \vJobAss{\idxJob}{\idxMach} = 1, \quad \symbJob_{\idxJob} \in \setJobs \label{eq:model-assignment} \\
    & \sum\limits_{\substack{\symbJob_{\idxJobAnother} \in \setJobsE\\\idxJobAnother \neq \idxJob}} \vJobPred{\idxJob}{\idxJobAnother}{\idxMach} = \vJobAss{\idxJob}{\idxMach}, \quad \symbJob_{\idxJob} \in \setJobs,\ \symbMach_{\idxMach} \in \setMach \label{eq:model-predecessor} \\
    & \sum\limits_{\substack{\symbJob_{\idxJobAnother} \in \setJobsS\\\idxJobAnother \neq \idxJob}} \vJobPred{\idxJobAnother}{\idxJob}{\idxMach} = \vJobAss{\idxJob}{\idxMach}, \quad \symbJob_{\idxJob} \in \setJobs,\ \symbMach_{\idxMach} \in \setMach \label{eq:model-successor} \\
    & \sum\limits_{\symbJob_{\idxJob} \in \setJobsE} \vJobPred{\idxJobAnother}{\idxJob}{\idxMach} = 1, \quad \symbMach_{\idxMach} \in \setMach, \ \symbJob_{\idxJobAnother} = \symbJobDS \label{eq:model-dummy-s} \\
    & \sum\limits_{\symbJob_{\idxJob} \in \setJobsS} \vJobPred{\idxJob}{\idxJobAnother}{\idxMach} = 1, \quad \symbMach_{\idxMach} \in \setMach, \ \symbJob_{\idxJobAnother} = \symbJobDE \label{eq:model-dummy-e} \\
    & \cR{\idxJob} \leq \vJobStart{\idxJob} \leq \cD{\idxJob} - \cP{\idxJob}, \quad \symbJob_{\idxJob} \in \setJobs \label{eq:model-rdp} \\
    \begin{split}
    \vJobStart{\idxJob} + \cP{\idxJob} + \vGap{\idxJob}{\idxJobAnother} \leq \vJobStart{\idxJobAnother} + \cBigM \cdot (1 - \vJobPred{\idxJob}{\idxJobAnother}{\idxMach}),\\ 
    \symbJob_{\idxJob} \in \setJobs,\ \symbJob_{\idxJobAnother} \in \setJobs,\ \symbMach_{\idxMach} \in \setMach  
    \end{split} \label{eq:model-overlaps} \\
    \begin{split}
    \vJobStart{\idxJobAnother} - (\vJobStart{\idxJob} + \cP{\idxJob}) \leq \vGap{\idxJob}{\idxJobAnother} + \cBigM \cdot (1 - \sum\limits_{\symbMach_{\idxMach} \in \setMach} \vJobPred{\idxJob}{\idxJobAnother}{\idxMach}),\\ 
    \symbJob_{\idxJob} \in \setJobs,\ \symbJob_{\idxJobAnother} \in \setJobs,\ \idxJob \neq \idxJobAnother
    \end{split} \label{eq:model-z-UB} \\
    \begin{split}
    \vJobStart{\idxJobAnother} - (\vJobStart{\idxJob} + \cP{\idxJob}) \geq \vGap{\idxJob}{\idxJobAnother} - \cBigM \cdot (1 - \sum\limits_{\symbMach_{\idxMach} \in \setMach} \vJobPred{\idxJob}{\idxJobAnother}{\idxMach}), \\ 
    \symbJob_{\idxJob} \in \setJobs,\ \symbJob_{\idxJobAnother} \in \setJobs,\ \idxJob \neq \idxJobAnother
    \end{split} \label{eq:model-z-LB} \\
    & \vGap{\idxJob}{\idxJobAnother} \leq \cBigM \cdot \sum\limits_{\symbMach_{\idxMach} \in \setMach} \vJobPred{\idxJob}{\idxJobAnother}{\idxMach}, \quad \symbJob_{\idxJob} \in \setJobs,\ \symbJob_{\idxJobAnother} \in \setJobs,\ \idxJob \neq \idxJobAnother \label{eq:model-z-zero} \\
    & \vMachOn{\idxMach} \geq (1-\vJobPred{\idxJob}{\idxJobAnother}{\idxMach}),\quad \symbMach_{\idxMach} \in \setMach, \ \symbJob_{\idxJob} = \symbJobDS, \ \symbJob_{\idxJobAnother} = \symbJobDE \label{eq:model-machine-on}
 \end{align}
 
 \endgroup
 
   Besides the constraints mentioned above, which define the behavior of the model, additional constraints can be added to reduce the search space by eliminating symmetries. Constraints \eqref{eq:model-sym-machine} state that jobs are preferably assigned to machines with lower indices. Constraints \eqref{eq:model-sym-task} pre-assign the first job to the first machine, the second job to the first or the second machine, etc. 
  
\begin{align}
    & \vMachOn{\idxMach} \geq \vMachOn{\idxMach + 1}, \quad \symbMach_{\idxMach} \in \{1,\dots,\cMach -1\} \label{eq:model-sym-machine} \\
    & \sum\limits_{\idxMach=1}^{\idxJob} \vJobAss{\idxJob}{\idxMach} = 1, \quad \idxJob \in \{1, \dots, \min \{ \cMach, \cJobs \} \} \label{eq:model-sym-task}
\end{align}

To further reduce the solution space, we use additional constraints that link the lengths of the gaps with the start times of the jobs -- stating that all the gaps and jobs should `fill' the whole scheduling horizon. For that, we add new variables $start_{\idxMach}$ and $end_{\idxMach}$, which denote the start time of the first job processed on machine $\symbMach_{\idxMach}$, and the ending time of the last job processed on machine $\symbMach_{\idxMach}$, respectively. It must hold that if job $\symbJob_{\idxJob}$ follows immediately after $\symbJobDS$ on machine $\symbMach_{\idxMach}$, then $start_{\cMach} = \vJobStart{\idxJob}$. Similarly, if job $\symbJob_{\idxJob}$ immediately precedes $\symbJobDE$ on machine $\symbMach_{\idxMach}$, then $end_{\idxMach} = \cHorizon - (\vJobStart{\idxJob} + \cP{\idxJob})$, where $\cHorizon$ is the length of the scheduling horizon. These logical implications in the form $\text{if} \ x = 1 \ \text{then} \ y = z$ with binary variable $x$ and continuous variables $y$ and $z$ can be linearized by introducing two constraints: 
\begin{equation*}
 y-z \leq \cBigM (1-x), \ \text{and} \ z-y \leq \cBigM (1-x).
\end{equation*}

\noindent Now, constraint \eqref{eq:model-sym-z} can be added:
   
\begin{align}
    & \sum_{\idxJob=1}^{\cJobs} \cP{\idxJob} + \sum_{\idxJob=1}^{\cJobs} \sum_{\substack{\idxJobAnother=1\\\idxJob \neq \idxJobAnother}}^{\cJobs} \vGap{\idxJob}{\idxJobAnother} + \sum_{\idxMach=1}^{\cMach} (start_{\idxMach} + end_{\idxMach}) = \cHorizon \sum_{\idxMach = 1}^{\cMach} \vMachOn{\idxMach}. \label{eq:model-sym-z}
\end{align}

\section{EXPERIMENTS} \label{sec:experiments}

\noindent To test the performance of the proposed model, we conduct two types of experiments. The first experiment compares our model to the position-based model adopted from the relevant paper \citep{2017_che}, and the second experiment examines the scalability of our model on larger problem instances.

For each experiment, a wide range of instances is generated for different combinations of $\cJobs$ and $\cMach$. The optimality gap is used to measure the performance of the model(s). In the following text, $\fUniform{x}{y}$ stands for an integer uniform distribution on interval $[x,y]$, $\fExponential{x}$ denotes the exponential distribution with scale parameter $x$, and $\fEx{\cP{\idxJob}}$ represents the expected processing time of job~$\idxJob$.

All experiments were performed on a Dell PC with an Intel Core i7-4610M CPU operating at 3 GHz, 16 GB RAM. Gurobi Optimizer (version 8.1) was used to solve the \abbrMILP{} models.

\subsection{Benchmark Data}

Jobs' parameters are generated according to the following scheme. At first, vector $\symbAssVec = (\symbAssElem{1}, \dots, \symbAssElem{\cJobs})$ is generated, \hbox{$\symbAssElem{\idxJob} \sim \fUniform{1}{\cMach}$}, describing the random assignment of the jobs to machines. Note that vector $\symbAssVec$ is used only for data generation, simulating the production process. It might be different from the assignment found by the optimization solver.  

Processing times, release times, and deadlines are generated according to (\ref{gen:processing}), (\ref{gen:release}) and (\ref{gen:deadline}), respectively. 

\begin{align}
    \cP{\idxJob} &\sim \fUniform{p_{min}}{p_{max}} \label{gen:processing} \\
    \cR{\idxJob} &\sim \sum_{k=1}^{\idxJob-1} \llbracket \symbAssElem{k} = \symbAssElem{\idxJob} \rrbracket \cdot \fEx{\cP{\idxJob}} + \fExponential{\alpha \cdot \fEx{\cP{\idxJob}}} \label{gen:release} \\
    \cD{\idxJob} &\sim \cR{\idxJob} + \cP{\idxJob} + \beta \cdot \fEx{\cP{\idxJob}} + \fExponential{\gamma \cdot \fEx{\cP{\idxJob}}} \label{gen:deadline}
\end{align}

Processing times, release times, and deadlines are assumed to be integers, so only the upper integer part of the generated data is considered. Symbols $p_{min}$, $p_{max}$, $\alpha$, $\beta$ and $\gamma$ represent parameters, which allow us to generate instances with different properties. Indicator $\llbracket \symbAssElem{k} = \symbAssElem{\idxJob} \rrbracket$ is one if $\symbAssElem{k} = \symbAssElem{\idxJob}$, and zero otherwise.

\subsection*{Experiment 1: Comparison with a Position-Based Model} \label{sec:comparison}

In this experiment, we compare our model to the position-based model, which was originally developed by \citet{2017_che} for a single machine problem. As the problem studied in this paper differs from the problem studied by Che et al., it was necessary to modify their model slightly. The modified model is described in Appendix. For both models, additional constraints (\ref{eq:model-sym-machine}--\ref{eq:model-sym-z}), and (\ref{eq:app-sb-1}--\ref{eq:app-sb-3}) were used.

Because the reference position-based model optimizes the energy consumption for two modes only, energy function $\fESymb_{600}$ with a single switching depicted in \Cref{fig:energy-furnace} was used for the experiment. There were 50 testing instances randomly generated for each combination of $\cJobs \in \{10,15,20\}$, $\cMach \in \{1,2,4\}$, $\alpha \in \{0.8, 1.2\}$ and $\gamma \in \{1, 1.5\}$; (1800 instances in total). Parameter $\beta$ was set to 1. The minimal and maximal processing times $p_{min}$, and $p_{max}$, were set to $1$ and $300$, respectively. The time limit was set to \secs{300} per instance.

\begin{figure}
    \centering
    \resizebox{\columnwidth}{!}{%
        \includegraphics{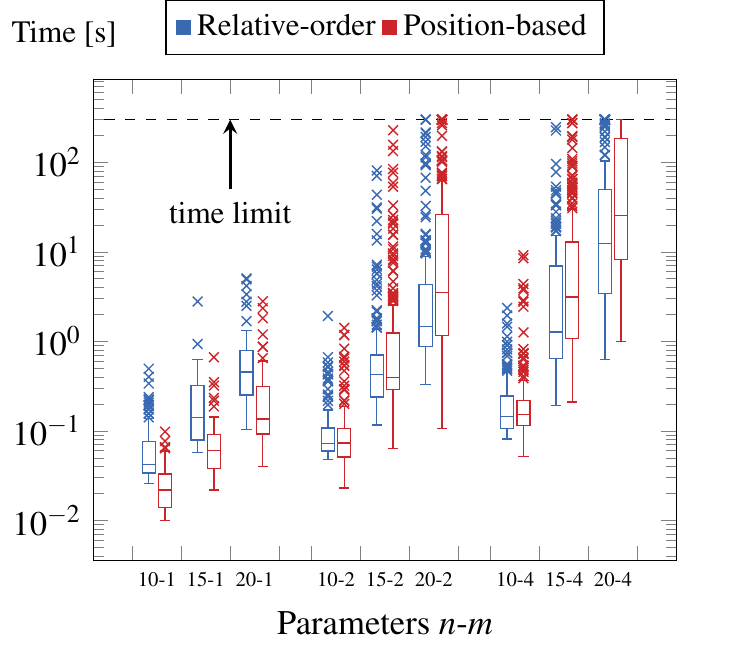}
    }
    \caption{Comparison between the relative-order model (left) and the position-based model (right) of the solving times on feasible instances for different pairs of $\cJobs$ and $\cMach$.}
    \label{fig:boxplot-time}
\end{figure}

The overall results of the experiment are shown in \Cref{tab:experiment-1-comparison-aggr}. Each row aggregates 200 instances; 50 generated for each combination of $\alpha$ and $\gamma$. A number of infeasible instances is given by \#if. A number of timeouts is listed in the table in the columns \#to. The average runtimes are measured and listed for feasible and infeasible instances separately, where $t_{\text{if}}$ and $t_{\text{f}}$ represent the average time over infeasible and feasible instances, respectively.
Average times that are typed in the bold font mark the better of the two tested solvers.
The average optimality gap is listed as well (computed over all instances aggregated in the given row). Solving times aggregated over all feasible instances are also depicted in \Cref{fig:boxplot-time} in the form of box plots.

\begin{table*}
    \vspace{10pt}
    \setlength{\tabcolsep}{3pt}
    \centering
    \caption{Comparison between the position-based model and the relative-order model.}
    \label{tab:experiment-1-comparison-aggr}
    \renewcommand*{\arraystretch}{1.0}
    \addtolength{\tabcolsep}{4pt}

    \begin{tabular}{ccr|rrrr|rrrr}  
    \toprule
    \multicolumn{3}{c}{Instances} &
    \multicolumn{4}{c}{Position-based (reference)} & 
    \multicolumn{4}{c}{Relative-order (this work)} \\
    \cmidrule(r){1-3} \cmidrule(r){4-7} \cmidrule(r){8-11}
    $\cJobs$ & $\cMach$ & \multicolumn{1}{c}{\#if} &
    \multicolumn{1}{r}{\#to} & $t_{\text{if}}$ [\secs{}] & $t_{\text{f}}$ [\secs{}] & \multicolumn{1}{r}{gap [\%]} &
    \multicolumn{1}{r}{\#to} & $t_{\text{if}}$ [\secs{}] & $t_{\text{f}}$ [\secs{}] & gap [\%] \\ \midrule
     & 1 &  87 &  0 &   \textbf{0.01} &  \textbf{0.03} & 0.00 &  0 &  0.03 &  0.08 & 0.00 \\
    10 & 2 &   6 &  0 &   \textbf{0.39} &  \textbf{0.12} & 0.00 &  0 &  0.41 &  \textbf{0.12} & 0.00 \\
     & 4 &   0 &  0 &   -    &  0.37 & 0.00 &  0 &  -    &  \textbf{0.24} & 0.00 \\ \midrule
     & 1 & 126 &  0 &   \textbf{0.02} &  \textbf{0.08} & 0.00 &  0 &  0.07 &  0.24 & 0.00 \\
    15 & 2 &  12 &  1 &  56.09 &  6.29 & 0.00 &  0 &  \textbf{1.99} &  \textbf{2.38} & 0.00 \\
     & 4 &   0 &  4 &   -    & 23.80 & 0.45 &  0 &  -    &  \textbf{8.64} & 0.00 \\ \midrule
     & 1 & 148 &  0 &   \textbf{0.04} &  \textbf{0.35} & 0.00 &  0 &  0.19 &  0.88 & 0.00 \\
    20 & 2 &  23 & 29 & 190.60 & 43.09 & 1.65 &  9 & \textbf{95.92} & \textbf{16.79} & 0.23 \\
     & 4 &   0 & 41 &   -    & 96.12 & 6.73 & 23 &  -    & \textbf{61.66} & 3.48 \\
    \bottomrule
    \end{tabular}
\end{table*}

The single machines instances were solved by both models without any problems. The performance of the position-based model was slightly better. However, the absolute times of both models were low. 
Instances with parallel machines are harder to solve because the assignment needs to be found together with the order of the jobs. Results show that our model outperformed the reference model when the number of machines was higher than 1. The biggest difference can be seen for \hbox{$\cMach = 2$}, \hbox{$\cJobs = 20$}, where the reference model ran out of time three times more often, and performed more than $2.5$ times slower on the feasible instances on average. The difference is not that large for $\cMach = 4$, $\cJobs = 20$, because both models started to reach the maximum solving time (see \Cref{fig:boxplot-time}).

Considering the optimality gap, the average overall instances for which at least one of the models did not find an optimal solution was \percents{23.02} for the reference model and \percents{9.84} for our model.

The total run time was \secs{37\,987} for the reference model and \secs{19\,894} for our model, so our model was nearly two times faster on average. Furthermore, it is important to note that our model can optimize any energy function, which can be approximated by a piecewise linear function, whereas the reference model was developed to optimize the on/off modes only.

\subsection*{Experiment 2: Scalability} \label{sec:larger-instances}

In the second experiment, the proposed model is tested on larger instances. For the experiment, a piecewise linear approximation of energy function $\fESymb_{\text{cont}}$ shown in \Cref{fig:ex-energy-func} was used. The approximation was made by 17 linear segments. Ten instances were generated for each combination of $\cJobs \in \{20, 25, 30, 35\}$, $\cMach \in \{2,4,6\}$, $\alpha \in \{1,1.5\}$, and $\gamma \in \{1,2\}$; 480 instances in total. Parameter $\beta$  was set to 1. Parameters $p_{min}$ and $p_{max}$ were again set to $1$ and $300$, respectively. The maximal time limit was set to \mins{10} per instance.

The results are listed in \Cref{tab:experiment-2}. Each row aggregates 40 instances generated for each combination of $\alpha$ and $\gamma$. Column \#if shows the number of infeasible instances, \#$\text{to}_{\text{if}}$ (\#$\text{to}_{\text{f}}$) represents the number of timeouts on infeasible (feasible) instances, $t_{\text{if}}$ ($t_{\text{f}}$) is the average time on the respective instances, and `gap' is the average optimality gap over the feasible instances, for which some solution was found.

\begin{table*}
    \vspace{10pt}
    \setlength{\tabcolsep}{3pt}
    \centering
    \caption{Performance of the relative-order model on larger instances.}
    \label{tab:experiment-2}
    \renewcommand*{\arraystretch}{1.0}
    \addtolength{\tabcolsep}{4pt}
    
    \begin{tabular}{ccr|rrrrr|rrrrr}  
    \toprule
    \multicolumn{3}{c}{Instances} &
    \multicolumn{5}{c}{Relative-order (without additional constr.)} & 
    \multicolumn{5}{c}{Relative-order (with additional constr.)} \\
    \cmidrule(r){1-3} \cmidrule(r){4-8} \cmidrule(r){9-13}
    $\cJobs$ & $\cMach$ & \multicolumn{1}{c}{\#if} & \multicolumn{1}{r}{\#$\text{to}_{\text{if}}$} &    \#$\text{to}_{\text{f}}$ & $t_{\text{if}}$ [\secs{}] & $t_{\text{f}}$ [\secs{}] & \multicolumn{1}{r}{gap [\%]} & \multicolumn{1}{r}{\#$\text{to}_{\text{if}}$} &    \#$\text{to}_{\text{f}}$ & $t_{\text{if}}$ [\secs{}] & $t_{\text{f}}$ [\secs{}] & \multicolumn{1}{r}{gap [\%]} \\ \midrule
       & 2 & 10 & 1 & 3  & 94  & 103 & 4.39  &	0  &	2  &	\textbf{56}  &	\textbf{56}  &	0.41 \\
    20 & 4 & 0  & 0 & 26 & -   & 452 & 36.07 &	0  &	11 &	-   &   \textbf{220} &	7.98 \\
       & 6 & 0  & 0 & 30 & -   & 501 & 47.17 &	0  &	13 &	-   &	\textbf{235} &	11.26 \\ \midrule
       & 2 & 4  & 2 & 10 & 305 & 229 & 7.22  &	1  &	5  &	\textbf{157} &	\textbf{141} &	2.27 \\
    25 & 4 & 0  & 0 & 33 & -   & 531 & 50.50 &	0  &	15 &	-   &	\textbf{270} &	12.08 \\
       & 6 & 0  & 0 & 39 & -   & 593 & 62.24 &	0  &	25 &	-   &	\textbf{432} &	19.18 \\ \midrule
       & 2 & 8  & 7 & 12 & \textbf{526} & 334 & 16.10 &	7  &	9  &	527 &	\textbf{220} &	5.15 \\
    30 & 4 & 0  & 0 & 38 & -   & 574 & 55.89 &	0  &    27 &	-   &	\textbf{464} &	23.44 \\
       & 6 & 0  & 0 & 40 & -   & 600 & 63.95 &	0  &	32 &	-   &	\textbf{518} &	25.99 \\ \midrule
       & 2 & 8  & 6 & 17 & 465 & 367 & 16.78 &	4  &	13 &	\textbf{363} &	\textbf{337} &	5.77 \\
    35 & 4 & 0  & 0 & 39 & -   & 592 & 54.65 &	0  &	24 &	-   &	\textbf{481} &	18.40 \\
       & 6 & 0  & 0 & 40 & -   & 600 & 65.95 &	0  &	38 &	-   &	\textbf{581} &	32.38 \\
    \bottomrule
\end{tabular}    

\end{table*}

\Cref{tab:experiment-2} shows how the complexity increases with the increasing number of jobs and machines. The performance of the model, which includes additional constraints \eqref{eq:model-sym-machine}, \eqref{eq:model-sym-task}, and \eqref{eq:model-sym-z}, is compared to the same model without these constraints. 

The results show that the additional constraints significantly improve the behavior of the model. The overall average optimality gap decreases from \percents{40.73} to \percents{14.20} when the constraints are used. Also, the number of optimally solved instances increases from 119 to 233, and the number of timeouts decreases from 343 to 226. The ability to find a feasible solution is comparable for both models -- model without symmetry breaking constraints finds a solution to 414 instances, while the model with symmetry breaking constraints finds a solution to 433 instances (out of 450 feasible instances).

When the instances are large, it may be hard to find a feasible solution or detect the infeasibility. However, this could be solved by a simple decomposition, using, for example, some heuristics or a Constraint programming model to check feasibility and possibly provide a feasible assignment as an initial solution to the \abbrMILP{} model.

\nopagebreak

\section{CONCLUSIONS} \label{sec:conclusions}

\noindent This study addresses the modeling of an idle energy minimization scheduling problems. The technique of implicit modeling of the machine modes called idle energy function, which abstracts dynamics of the machine and provides a link between the idle period length and the optimal idle energy consumption, is adopted from the domain of embedded systems. It is shown that this method is applicable to a wide range of idle energy minimization problems. Furthermore, discussed examples illustrate that the properties of the problems are similar across different domains, and the shape of the energy function is the same for many relevant applications.

An efficient \abbrMILP{} model that uses the idle energy function is proposed to solve the scheduling problem optimally. The proposed model is compared to the position-based model adapted from the literature. The experiment shows that the overall performance of our model is significantly better when the number of machines was higher than one, even though the reference model is less general. Besides the comparison, another experiment is conducted to show the performance of the proposed model and the effect of the symmetry breaking constraints on larger instances of the problem. The importance of the additional constraints is apparent, as the overall optimality gap decreases nearly three times when the constraints are used.

In our future research, we would like to integrate the idle energy function into heuristics to solve industrial-size instances of the problem. Also, we want to investigate multi-objective optimization, because the trade-off between energy consumption and productivity-related objectives (such as makespan or the total tardiness) is known and widely studied.


\section*{\uppercase{Acknowledgements}}

\noindent This work was funded by the Ministry of Education, Youth and Sport of the Czech Republic within the project Cluster 4.0 number CZ.02.1.01/0.0/0.0/16\_026/0008432. This work was also supported by the Grant Agency of the Czech Technical University in Prague, grant No. SGS19/175/OHK3/3T/13.

{\small
\bibliography{bibliography}}

\newpage

\section*{\uppercase{Appendix}}

\noindent Here, the position-based \abbrMILP{} model used for the comparison is described. It was originally proposed in \citep{2017_che} to minimize the total tardiness and idle energy on a single machine with a single power-saving mode.

\subsection*{Reference Model}

The idea of the model is to represent all possible positions to which the individual jobs can be assigned. The variable representing the completion time is linked with the position instead of the job. A set of constraints assure that if a job is assigned to some position, its completion time is bounded (by the deadline, neighboring jobs, etc.). Following decision variables are used:

\begin{itemize}[leftmargin=2.75em]
  \item[$x_{i,l,k}$] Binary variable; if job $\symbJob_{i}$ is assigned to  position $l$ on machine $\symbMach_{k}$. then $x_{i,l,k} = 1$, otherwise 0
  \item[$y_{l,k}$] Binary variable; if there is turn-off-on operation immediately after $l$-th job is processed on machine $\symbMach_{k}$, then $y_{l,k} = 1$, otherwise 0
  \item[$c_{l,k}$] Integer variable; completion time of $l$-th job on machine $\symbMach_{k}$
  \item[$E_{l,k}$] Continuous variable; energy consumed by machine $\symbMach_{k}$ between the completion time of $k$-th job and start of $(k+1)$-th job
\end{itemize}

\noindent To simplify the notation, we substitute $p_{i,l,k}$ for $\sum_{i=1}^{\cJobs} x_{i,l,k} \cdot \cP{i}$. Also, we substitute $\mathcal{L}$ for $\{1,2, \dots, \cJobs\}$. In the following model, $T_{sw}$ denotes the switching time (i.e., the break-even time between the operating and standby modes), and $C_{sw}$ stands for the switching cost; $P_{on}$ and $P_{sb}$ represent the power consumed in the processing mode and the standby mode, respectively. Now, the whole model can be written as follows:

\begingroup
\allowdisplaybreaks

\begin{align}    
& \min \sum\limits_{k=1}^{\cMach} \sum\limits_{l=1}^{\cJobs-1} E_{l,k} \\
& \text{subject to} \nonumber \\
& \sum\limits_{l=1}^{\cJobs} \sum\limits_{k=1}^{\cMach} x_{i,l,k} = 1, \quad \symbJob_{i} \in \setJobs \\
& \sum\limits_{i=1}^{\cJobs} x_{i,l,k} \leq 1, \ l \in \mathcal{L}, \quad \symbMach_{k} \in \setMach \\
& c_{l,k} - p_{i,l,k} \geq \sum\limits_{i=1}^{\cJobs} x_{i,l,k} \cdot \cR{i}, \quad l \in \mathcal{L}, \ \symbMach_{k} \in \setMach \\
\begin{split}
	c_{l,k} \leq \sum\limits_{i=1}^{\cJobs} \left(x_{i,l,k} \cdot \cD{i} \right) + \cBigM \cdot \left(1-\sum\limits_{i=1}^{\cJobs} x_{i,l,k}\right), \\ l \in \mathcal{L}, \  \symbMach_{k} \in \setMach
\end{split} \\
& c_{l,k} \leq c_{l-1,k} + p_{i,l,k} + y_{l,k} \cdot T_{sw}, \quad l \in \mathcal{L}, \ \symbMach_{k} \in \setMach \\
\begin{split}
	E_{l,k} \geq  ( c_{l+1,k} - c_{l,k} - p_{i,l,k} ) \cdot P_{on} - \cBigM \cdot y_{l,k}, \\ l \in \mathcal{L}, \ \symbMach_{k} \in \setMach
\end{split} \\
\begin{split}
    E_{l,k} \geq C_{sw} \cdot y_{l,k} + (c_{l+1,m} - c_{l,m} - p_{i,l,k} - T_{sw}) \cdot P_{sb}, \\
    l \in \mathcal{L}, \ \symbMach_{k} \in \setMach
\end{split} \label{eq:app-standby}
\end{align}
\endgroup

To improve the performance of the model, we add several symmetry-breaking constraints. Constraints \eqref{eq:app-sb-1} and \eqref{eq:app-sb-2} are analogous to constraints \eqref{eq:model-sym-machine} and \eqref{eq:model-sym-task}, respectively. Constraint \eqref{eq:model-sym-z} cannot be easily integrated into the described position-based model as it does not use the variables $\vGap{\idxJob}{\idxJobAnother}$. Instead, constraints \eqref{eq:app-sb-3} are added, which enforce assignment order from the leftmost position to the right.

\begin{align}
    & \sum_{i=1}^{\cJobs} x_{i,1,k} \geq \sum_{i=1}^{\cJobs} x_{i,1,k+1}, \quad k \in \{1,\dots,\cMach-1\} \label{eq:app-sb-1} \\ 
    & \sum_{l=1}^{\cJobs} \sum_{k=1}^{i} x_{i,l,k} = 1, \quad i \in \{1, \dots, \min \{ \cMach, \cJobs \} \} \label{eq:app-sb-2} \\
    & \sum_{i=1}^{\cJobs} x_{i,l,k} \geq x_{i,l+1,k}, \quad k \in \{1,\dots, \cJobs-1\}, \ \symbMach_{k} \in \setMach \label{eq:app-sb-3}
\end{align}

Structure of the constraints is the same as proposed originally by Che et al. For a detailed description, we refer the reader to the original publication \citep{2017_che}. The main adjustments, which were made to fit our problem statements are: (i) the original variables $x_{i,l}$ modeling the assignment were extended to $x_{i,l,k}$; (ii) original tardinesses were omitted and replaced by hard deadlines; (iii) constraint \eqref{eq:app-standby} was slightly changed to work even for non-zero standby power; (iv) several symmetry breaking constraints were added to improve the performance. The modified position-based model contains $\mathcal{O}(\cMach \cdot \cJobs^2)$ variables, which is asymptotically comparable to our relative-order model.

\end{document}